\documentstyle[prd,aps,amssymb,eqsecnum,twocolumn,floats]{revtex}

\def\beq{\begin{equation}}
\def\eeq{\end{equation}}
\def\beqa{\begin{eqnarray}}
\def\eeqa{\end{eqnarray}}
\def\bfig{\begin{figure}}
\def\efig{\end{figure}}

\input{psfig}
\begin{document}
\fnsymbol{footnote}
\draft
\wideabs{

\title{Stability of the $r$-modes in white dwarf stars}

\author{Lee Lindblom}
\address{Theoretical Astrophysics 130-33,
         California Institute of Technology,
         Pasadena, CA 91125}

\date{\today}
\maketitle

\begin{abstract}
Stability of the $r$-modes in rapidly rotating white dwarf stars is
investigated.  Improved estimates of the growth times of the
gravitational-radiation driven instability in the $r$-modes of the
observed DQ Her objects are found to be longer (probably considerably
longer) than $6\times 10^9$y.  This rules out the possibility that the
$r$-modes in these objects are emitting gravitational radiation at
levels that could be detectable by LISA.  More generally it is shown
that the $r$-mode instability can only be excited in a very small
subset of very hot ($T\gtrsim 10^6$K), rather massive ($M\gtrsim
0.9M_\odot$) and very rapidly rotating ($P_{\min}\leq P\lesssim
1.2P_{\min}$) white dwarf stars.  Further, the growth times of this
instability are so long that these conditions must persist for a very
long time ($t\gtrsim 10^9$y) to allow the amplitude to grow to a
dynamically significant level.  This makes it extremely unlikely that
the $r$-mode instability plays a significant role in any real white
dwarf stars.
\end{abstract}

}

\section{Introduction}
\label{sectionI}
Recently it was discovered by Andersson~\cite{nils}, and Friedman and
Morsink~\cite{friedman-morsink} that gravitational radiation tends to
drive unstable the $r$-modes of all rotating stars.  Lindblom, Owen,
and Morsink~\cite{lom} subsequently calculated the strength of the
gravitational radiation coupling to these modes and found it to be
strong enough to overcome viscous dissipation in hot rapidly rotating
neutron stars.  This calculation was confirmed using better estimates
of the dissipative coupling by Andersson, Kokkotas, and
Schutz~\cite{aks} and Lindblom, Mendell, and Owen~\cite{lmo}.  Thus it
is now generally expected that this $r$-mode instability will limit
the rotation rates of hot young rapidly rotating neutron stars.  The
excess angular momentum in these stars will be carried away by
gravitational radiation, and these stars will be spun down to more
slowly rotating stable configurations within about one year of their
births.  The gravitational radiation emitted during this process may
eventually be detectable by LIGO, (see Owen, et al.~\cite{owenetal},
and Brady and Creighton~\cite{bc}).

The possibility that this $r$-mode instability plays a role in other
astrophysical systems has also been proposed.  For example, the
possibility that this instability might play a role in limiting the
angular velocities of old and relatively cold neutron stars spun up by
accretion was considered by Bildsten~\cite{bildsten}, Andersson,
Kokkotas, and Stergioulas~\cite{akst}, and Levin~\cite{levin}.  It has
also been proposed that this instability might play a role in rapidly
rotating white dwarf stars by Andersson, Kokkotas, and
Stergioulas~\cite{akst} and by Hiscock~\cite{hiscock}.  This last
possibility will be investigated more thoroughly in this paper.  A set
of rapidly rotating white dwarf models, and the timescales for the
gravitational-radiation driven $r$-mode instability in these stars are
computed in Sec.~\ref{sectionII}.  The possibility that this
instability is playing a significant role in the observed DQ Her
objects is examined in Sec.~\ref{sectionIII}.  It is shown that the
minimum growth time for this instability in these objects is $6\times
10^9$y, and that the actual growth time is almost certainly much
longer than this.  Thus, the $r$-mode instability is not playing any
significant role in these objects, and gravitational radiation from the
$r$-modes in these objects will not be observable by LISA.
Sec.~\ref{sectionIV} evaluates the effects of viscosity on the
$r$-modes in white dwarf stars.  It is shown that all white dwarfs
with core temperatures cooler than about $10^6$K are stable; that all
white dwarfs with masses less than $0.9M_\odot$ are stable; and that
all white dwarfs with rotation periods longer than $1.2P_{\min}$
(where $P_{\min}$ is the minimum rotation period of that star) are
stable.  It is shown that white dwarf stars cool too quickly (in the
absence of accretion) for the $r$-mode instability to grow to
significant levels in any star.  And finally, it is shown that this
instability can only play a significant role in white dwarf stars that
are maintained at very high core temperatures by accretion, and that
remain at nearly maximal rotation rates for a period of time that
exceeds about $10^9$y.  It seems very unlikely that these conditions
will ever be met in any real white dwarf stars.

\section{Rapidly Rotating White Dwarfs}
\label{sectionII}

In order to study to properties of the $r$-modes in white dwarf stars,
simple models have been constructed using the equation of state of a
zero-temperature degenerate Fermi gas with $Z/A=\case{1}{2}$ that is
appropriate for carbon-oxygen white dwarfs~\cite{st}.  Families of
rigidly rotating stars were constructed using the numerical techniques
described by Ipser and Lindblom~\cite{ipser-lind} for finding very
rapidly rotating and highly non-spherical models.  A constant mass
family of rotating models was constructed by successively spinning up
a non-rotating model until its angular velocity was equal (or very
nearly equal) to the frequency of the equatorial orbit located just at
the surface of the star.  In Fig.~\ref{fig1} are presented the
rotation periods $P_{\min}$ of these maximally rotating white dwarf
models as a function of the mass of the star.  Also shown in
Fig.~\ref{fig1} are a set of points that represent the minimum
rotation periods for white dwarf stars as determined by
Hachisu~\cite{hachisu}.  The numerical models constructed here used a
spherical grid that allowed a much finer determination of the location
of the surface of the star than the models constructed by Hachisu.
Nevertheless, there is good agreement between our calculations of
the limiting rotation periods of these stars.

\bfig \centerline{\psfig{file=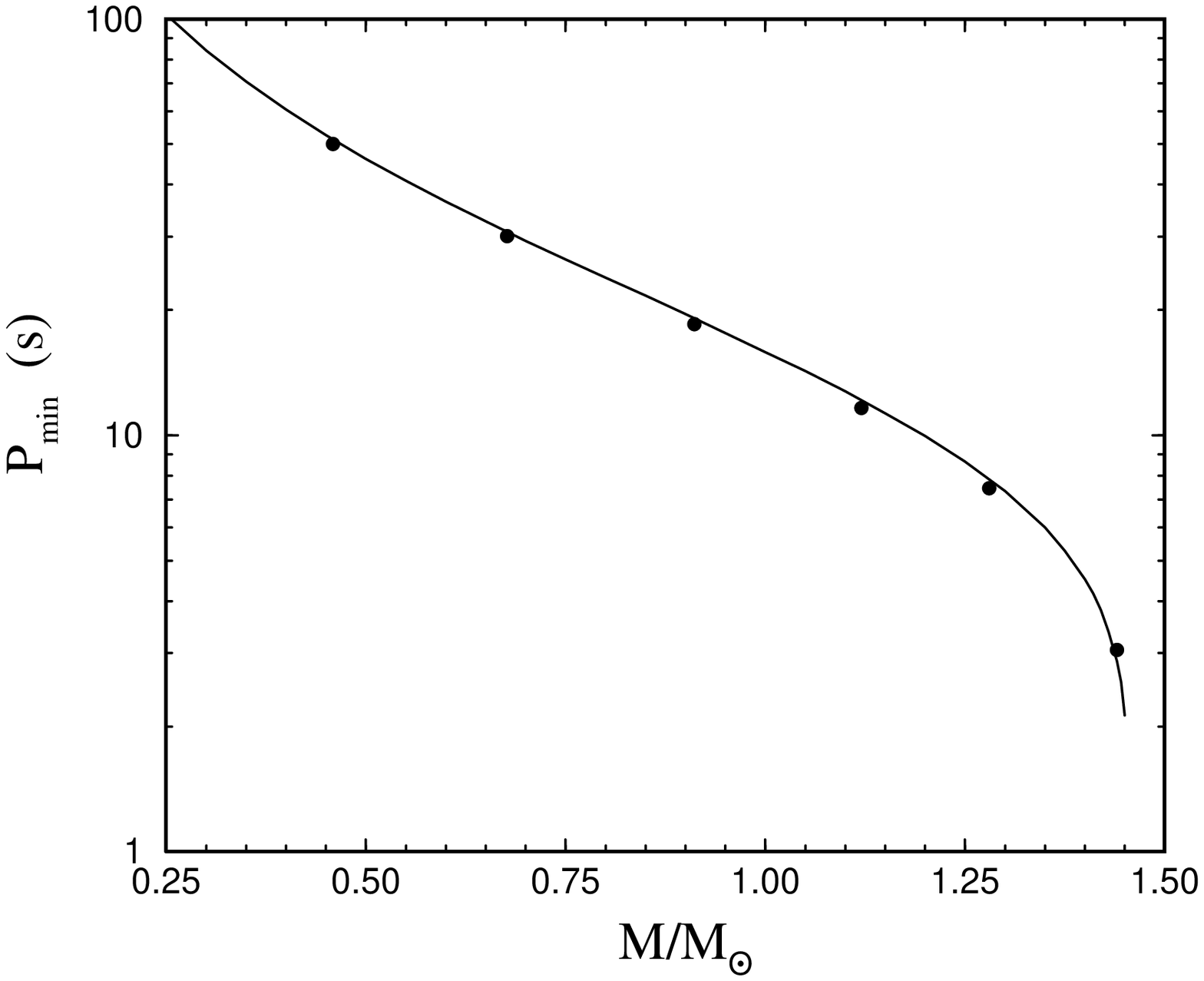,height=2.4in}} \vskip 0.3cm
\caption{Minimum rotation periods (in seconds) for rigidly rotating
white dwarf stars based on the equation of state of a
zero-temperature degenerate Fermi gas with
$Z/A=\case{1}{2}$.\label{fig1}} \efig

The growth time of the gravitational-radiation driven $r$-mode
instability is estimated using the expressions derived by Lindblom,
Owen, and Morsink~\cite{lom}.  In particular the growth time
$\tau_{\scriptscriptstyle GR}$ for the dominant $m=2$ mode is

\beq {1\over \tau_{\scriptscriptstyle GR}}= {2\pi\over 25}\left({4\over
3}\right)^8 {G\over c} \int_0^R \rho
\left({r\Omega\over c}\right)^6dr,\label{2.1} \eeq

\noindent where $\rho$ and $\Omega$ are the density and angular
velocity of the equilibrium stellar model, and $G$ and $c$ are
Newton's constant and the speed of light respectively.  This is only
the lowest order term in the expansion for $1/\tau_{\scriptscriptstyle
GR}$ in powers of the angular velocity.  To lowest order therefore it
is sufficient to evaluate this integral using the density function
$\rho$ for the non-rotating model of a given mass.  The gravitational
radiation growth time $\tau_{\scriptscriptstyle GR}$ defined in
Eq.~(\ref{2.1}) is proportional to $\Omega^{-6}$.  It is convenient to
define a characteristic growth time $\tilde{\tau}_{\scriptscriptstyle
GR}$ therefore for the star rotating at the maximum possible angular
velocity.  In particular then the gravitational radiation growth time
is related to this characteristic growth time by

\beq
\tau_{\scriptscriptstyle GR} = \tilde{\tau}_{\scriptscriptstyle GR}
\left({P\over P_{\min}}\right)^6.\label{2.2}
\eeq

\noindent The values of the characteristic growth time
$\tilde{\tau}_{\scriptscriptstyle GR}$ for a range of white dwarf
masses are presented in Fig.~\ref{fig2}.

\bfig \centerline{\psfig{file=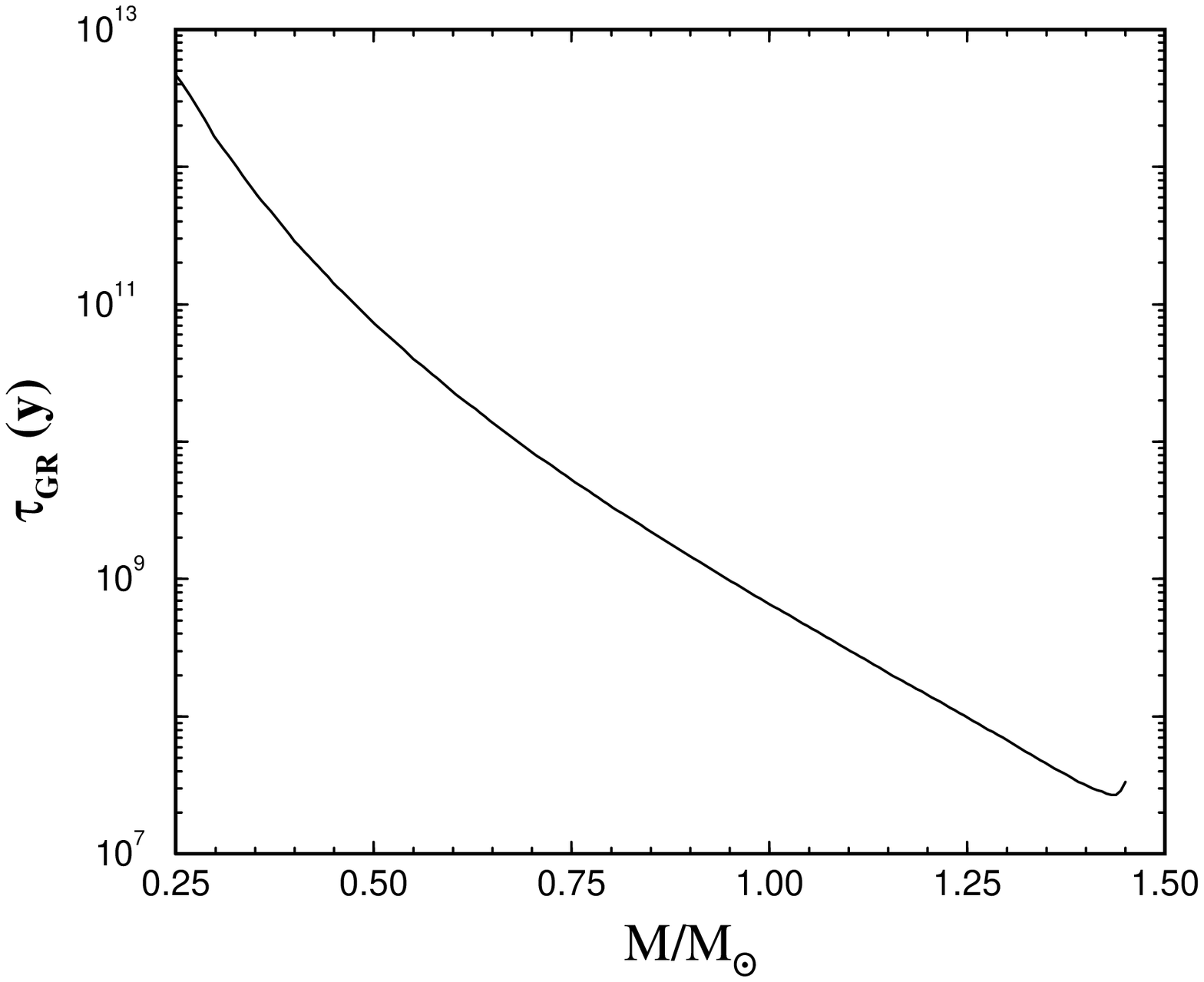,height=2.4in}} \vskip 0.3cm
\caption{Gravitational radiation growth times (in years) of the $r$-modes
for white dwarf stars rotating at their maximum possible rotation
rates.\label{fig2}} \efig

\section{DQ Her Objects}
\label{sectionIII}

Recently Andersson, Kokkotas, and Stergioulas~\cite{akst} and
Hiscock~\cite{hiscock} have suggested that the instability of the
$r$-modes driven by gravitational radiation may be playing an
important astrophysical role in the DQ Her type cataclysmic variable
systems.  These objects emit radiation with periodic luminosity
variations that have been interpreted as the rotation periods of
rapidly rotating white dwarf stars.  The systems having the shortest
periods are WZ Sge with period 27.9s, AE Aqr with 33.1s, V533 Her with
63.6s, and DQ Her with 71.1s~\cite{patterson}.  If these represent the
rotation periods of the white dwarfs (an interpretation that is not
universally accepted~\cite{zrsh,cssh}) then Fig.~\ref{fig1} shows that
these systems could (depending on their masses) contain very rapidly
rotating white dwarf stars.  The gravitational radiation growth times
of the $r$-modes in these systems are smallest---for a star of given
mass---for the most rapidly rotating stars.  It can also be shown (by
brute force numerical examination of the data in Fig.~\ref{fig2}
together with Eq.~\ref{2.2}) that the gravitational radiation growth
times of the $r$-modes in these systems are also smallest---for stars
of given rotation period---for that star which has the smallest mass.
Hence it is straightforward to show from the data in Figs.~\ref{fig1}
and \ref{fig2} that the minimum possible gravitational radiation
growth times for the $r$-modes in the DQ Her objects are $6.7\times
10^9$y for WZ Sge, $1.5\times 10^{10}$y for AE Aqr, $3.7\times
10^{11}$y for V533 Her, and $6.6\times 10^{11}$y for DQ Her.

The estimates of the gravitational radiation growth times given above
are almost certainly significant underestimates.  First, internal
fluid dissipation (i.e. viscosity) has been completely neglected.
This will always act to increase these timescales.  Second, the
assumption that these are maximally rotating white dwarf stars is
quite questionable and consequently the gravitational radiation growth
times are expected to be much longer than the values given above.  For
example, the most recent mass determination of the white dwarf in WZ
Sge is only $0.3 M_\odot$~\cite{cssh}.\footnote{A white dwarf with
rotation period 27.9s must have at minimum a mass of 0.725$M_\odot$.
Therefore, if the rotation period of WZ Sge is 27.9s, then its mass
must be greater than 0.725$M_\odot$.  Conversely, if its mass is less
than 0.725$M_\odot$ as suggested by recent mass determinations, then
the observed 27.9s period must be a harmonic of its fundamental
rotation period, or be caused by a pulsation or other phenomenon
unrelated to its rotation.}  The minimum gravitational
radiation growth time for a star of mass $0.3 M_\odot$ is $1.6\times
10^{12}$y.  The most recently measured mass for AE Aqr is
$0.8M_\odot$~\cite{cmmh}, and for DQ Her is $0.6M_\odot$~\cite{hww}.
These values imply that these objects are rotating significantly below
their maximum rotation rates: ${P/P_{\min}} = 1.4$ for AE Aqr and
${P/P_{\min}} = 2.0$ for DQ Her.  The gravitational radiation growth
times for these systems would then be $2.4\times 10^{10}$y for AE Aqr
and $1.3\times 10^{12}$y for DQ Her.  And further, recent observations
indicate that the rotation period of the white dwarf in DQ Her is
142.2s rather than 71.1s~\cite{zrsh}.  This increases the
gravitational growth time for this star to $8.1\times 10^{13}$y.

In order for the instability in the $r$-modes to play a significant
dynamical role in the evolution of a system, the dimensionless
amplitude of the mode must grow to a value of order unity.  For
example, the gravitational radiation amplitudes computed by
Hiscock~\cite{hiscock} are based on a presumed balance between
accretion and gravitational-radiation torques on the star.  Using the
equations in Owen {\it et al.}~\cite{owenetal}, it is easy to show
that (in a rapidly rotating star) this balance requires the
dimensionless amplitude of the $r$-mode to be a constant of order
unity multiplied by $(\tau_{\scriptscriptstyle GR} \dot{M}/M)^{1/2}$.
Given the observed accretion rates, the amplitudes of the $r$-modes
would have to be of order unity in these systems to maintain this
balance.  The extreme lengths of the gravitational-radiation
timescales in these systems means that there is not enough time for
the amplitude of the $r$-mode to grow to such a large value.  So even
if viscosity were unimportant and all white dwarfs were unstable to
the gravitational-radiation driven $r$-mode instability, this
instability can not be playing any significant role in the observed DQ
Her objects.  Gravitational radiation from the $r$-mode instability in
these objects will not be detectable by LISA.

\section{Viscous Dissipation}
\label{sectionIV}

Shear viscosity tends to suppress the gravitational radiation driven
instability in the $r$-modes.  In particular the growth time $\tau$ of the
mode becomes

\beq {1\over \tau} = {1\over\tau_{\scriptscriptstyle GR}} - {1\over
\tau_{\scriptscriptstyle V}},\label{4.0} \eeq

\noindent where $\tau_{\scriptscriptstyle V}$ is the viscous
timescale.  For the $m=2$ $r$-mode of primary interest to us here,
$\tau_{\scriptscriptstyle V}$ is given by the expression~\cite{lom}

\beq 
\tau_{\scriptscriptstyle V} 
= \case{1}{5} \int_0^R \rho r^6 dr \left(\int_0^R \eta r^4
dr\right)^{-1},\label{4.1}
\eeq

\noindent where $\eta$ is the shear viscosity.  The dominant form of
shear viscosity in hot white dwarf stars is expected to be 
electron scattering with the ion liquid.  Nandkumar and
Pethick~\cite{NP} provide the following analytical fit for the
shear viscosity in a ${}^{12}C$ ion liquid,

\beq
\eta = {10^6\rho_6^{2/3}\over \Bigl(1+1.62\rho_6^{2/3}\Bigr)I_2}
\label{4.2}
\eeq

\noindent where $I_2$ is

\beqa
I_2 = &&0.667 \log\Bigl(1.32+0.103T_6^{1/2}\rho_6^{-1/6}\Bigr)\nonumber\\
&&\quad+0.611 - {0.475+1.12\rho_6^{2/3}\over 1+1.62\rho_6^{2/3}}.
\label{4.3}
\eeqa

\noindent The density $\rho_6$ and temperature $T_6$ are to be given
here in units of $10^6$gm/cm${}^3$ and $10^6$K respectively.  This
formula for $\eta$ is sufficiently accurate (i.e. to within about 10\%)
for the densities above $10^6$gm/cm${}^3$ which dominate the integral
in the denominator of Eq.~(\ref{4.1}).

The growth time $\tau$ defined in Eq.~(\ref{4.0}) is negative for
slowly rotating stars: $P \gg P_{\min}$.  In this case viscosity
dominates and the $r$-mode instability is suppressed.  For more
rapidly rotating stars, however, this expression may become positive
and hence the $r$-modes unstable.  The critical rotation period
$P_c$ where the instability first sets in is determined
by setting $1/\tau=0$.  It follows then that

\beq
{P_{\min}\over P_c}=\left({\tilde{\tau}_{GR}\over 
\tau_{\scriptscriptstyle V}}\right)^{1/6}.
\label{4.4}
\eeq

\noindent This critical rotation period $P_c$ has been evaluated for
the white dwarf models discussed in Sec.~\ref{sectionII} and a range
of core temperatures appropriate for hot white-dwarf stars.  These
results are depicted graphically in Fig.~\ref{fig3}.  These results
show that no white dwarf is unstable to the $r$-mode instability when
its core temperature falls below $10^6$K.  Further, no white dwarf
star with mass smaller than 0.9$M_\odot$ is subject to the $r$-mode
instability if its core temperature is below $2\times 10^8$K, the
maximum core temperature expected to occur in accreting white dwarf
systems~\cite{nty,ni}.  No white dwarf with rotation period greater
than 1.2$P_{\min}$ is unstable to the $r$-mode instability if its core
temperature is below $2\times 10^8$K.

 \bfig \centerline{\psfig{file=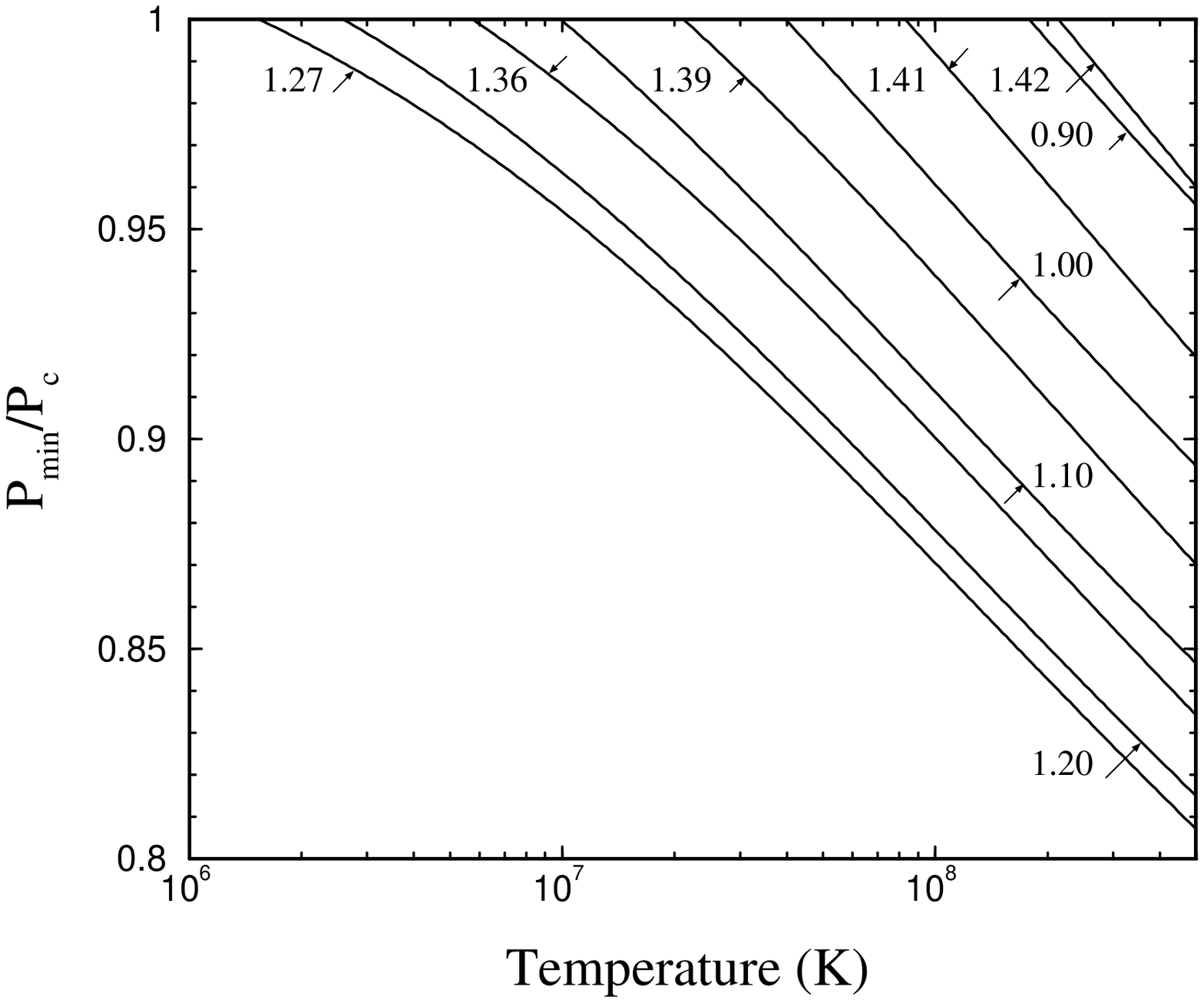,height=2.4in}} \vskip 0.3cm
\caption{Critical rotation periods $P_c$ where the gravitational
radiation instability in the $r$-modes first sets in: stars with
$P<P_c$ are unstable.  The mass of the white dwarf (in units of
$M_\odot$) labels each curve.\label{fig3}} \efig

To investigate further whether the $r$-mode instability might be
playing a role in some white dwarf stars, the growth times $\tau$ for
these modes are examined.  The expression for this timescale,
Eq.~(\ref{4.0}), can be re-written in terms of the quantities graphed
in Figs.~\ref{fig1}, \ref{fig2} and \ref{fig3}:

\beq
\tau = \tilde{\tau}_{GR}\left[\left({P_{\min}\over P}\right)^6
- \left({P_{\min}\over P_c}\right)^6\right]^{-1}.\label{4.5}
\eeq

\noindent The timescale $\tau$ becomes infinite as $P\rightarrow P_c$
and is minimum for $P=P_{\min}$.  Thus it is useful to define the
smallest possible growth time $\tau_{\min}$ for a star of given
mass and temperature:

\beq
\tau_{\min} = \tilde{\tau}_{GR}\left[1-\left({P_{\min}\over P_c}\right)^6
\right]^{-1}.\label{4.6}
\eeq

 \bfig \centerline{\psfig{file=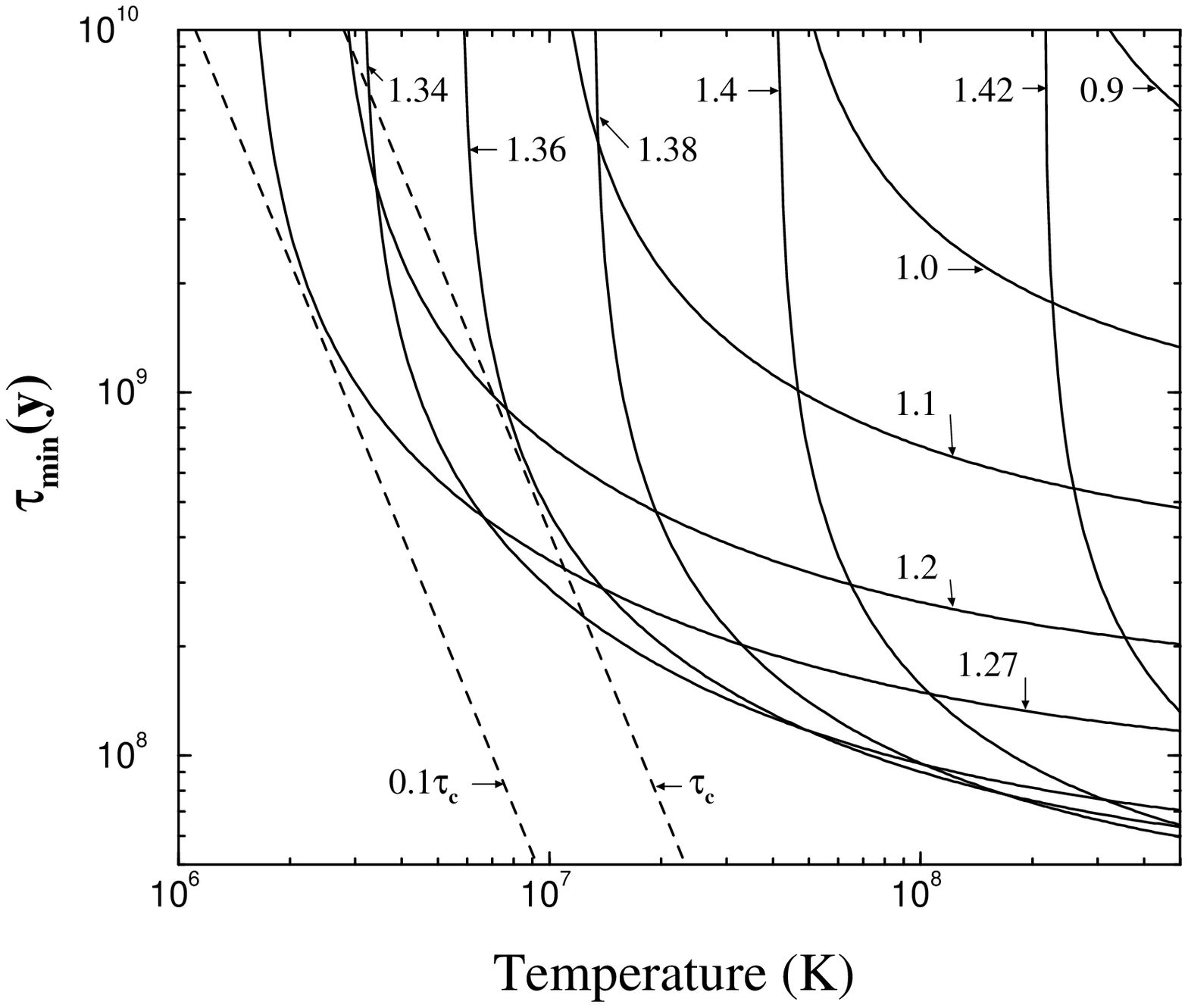,height=2.4in}} \vskip 0.3cm
\caption{Minimum growth times $\tau_{\min}$ for the gravitational
radiation driven instability of the $r$-modes for white dwarf
stars. The mass of the star (in units of $M_\odot$) labels each curve.
The dashed curves depict the time $\tau_c$ needed for a white dwarf to
cool to a given core temperature.
\label{fig4}} \efig

\noindent The values of these minimum growth times are depicted in
Fig.~\ref{fig4}.  It follows that no white dwarf has an $r$-mode
instability with growth time shorter than $7\times 10^7$ years if its core
temperature is below $2\times 10^8$K.  

In the absence of accretion a white dwarf star will quickly cool.  Its
core temperature drops to the value $T$ in approximately the time
$\tau_c$ given by the expression~\cite{st2},

\beq
\tau_c \approx 1.3\times 10^{11} T_6^{-5/2} {\rm y}.\label{4.7}
\eeq

\noindent The dashed curves in Fig.~\ref{fig4} represent this cooling
time $\tau_c$ and also $0.1\tau_c$.  It follows that most of the minimum
$r$-mode instability growth times are longer than $\tau_c$.  In these
stars, the amplitude of the $r$-mode will grow by less than a factor
of $e$ before the star cools and the instability is suppressed.  No
star has a minimum instability growth time that is shorter than
$0.1\tau_c$.  Thus, the $r$-mode instability will never have time to
grow to a dynamically important level in any white dwarf except during
the time it is heated by accretion.

If a white dwarf is heated by accretion, this heating can last for
only a finite time, given approximately by $\tau_{\scriptscriptstyle
A} = \Delta M/\dot{M}$ where $\Delta M$ is the total amount of
accreted material and $\dot{M}$ is the accretion rate.  Since the
$r$-mode instability is never effective for stars with $M< 0.9M_\odot$
we see that the $r$-mode instability can only act for a time when
$\Delta M<0.6M_\odot$.  Further, we see from Fig.~\ref{fig4} that
$\tau_{\min}\gtrsim 7\times 10^7$y for stars with $T\lesssim 2\times
10^8$K, the maximum core temperature reached in models of accretion
onto white dwarfs~\cite{nty,ni}.  In order for the $r$-mode
instability to have time to grow to a dynamically significant level
then, it would be necessary to have $\tau_{\scriptscriptstyle
A}\gtrsim10\tau_{\min}$.  This implies that the accretion rate must
satisfy $\dot{M}\lesssim 9\times 10^{-10} M_\odot y^{-1}$.  Given this
low accretion rate, this argument can be strengthened slightly.  For
white dwarfs with $\dot{M}\lesssim 10^{-9}M_\odot y^{-1}$ the core
temperature is only expected to reach about $8\times
10^7$K~\cite{nty,nI} during accretion and so from Fig.~\ref{fig4}
$\tau_{\min}\gtrsim 10^8$y.  For stars with core temperatures in this
range, Fig.~\ref{fig4} implies that instability can only occur on a
timescale fast enough to act effectively within the age of the
universe for stars in the mass range $1.05\lesssim M \lesssim 1.4
M_\odot$ and so $\Delta M\lesssim 0.35M_\odot$.  Thus, the upper limit
on the accretion rate can be improved slightly to $\dot{M}\lesssim
3.5\times 10^{-10}M_\odot y^{-1}$.

One further difficulty to achieving a dynamically significant $r$-mode
instability in white dwarf stars is illustrated in Fig.~\ref{fig5}.
The minimum growth times $\tau_{\min}$ shown in Fig.~\ref{fig4} only
pertain to stars that are maximally rotating $P=P_{\min}$.  The actual
growth times of these modes increase rapidly for more slowly rotating
stars, as illustrated in Fig.~\ref{fig5} for the $1.3M_\odot$ stellar
model.  For example, the growth time increases from about $10^8$y for
the maximally rotating model with $T=2\times 10^8$K to about $10^9$y
for the still very rapidly rotating model with $P=1.15P_{\min}$.  Thus
the very hot ($T\gtrsim3\times 10^6$K) and very rapid-rotation
conditions ($P_{\min}\leq P \lesssim 1.15P_{\min}$) needed to allow
the growth of the $r$-mode, must be maintained for a very long time.
This can only be achieved by a very low and steady accretion rate that
adds (or removes) angular momentum from the star at just the rate
needed to maintain maximal rotation as the star's mass
increases.\footnote{The angular momentum of a maximally rotating white
dwarf star decreases as its mass increases for stars with $M\gtrsim
1.1M_\odot$~\cite{hachisu}.}  In the most favorable case (i.e. by
keeping temperatures above $10^8$K, rotation periods less than
$1.05P_{\min}$ and the mass of the star near $1.3M_\odot$) these
special conditions need last only about $10^9$y in order to allow the
amplitude of the mode to grow by a factor of $e^{10}\approx 2\times
10^4$.  In the more typical and less favorable cases this type of
accretion would be needed for $10^{10}$y or longer to achieve this
amplification of the mode.  It seems rather unlikely that these
special conditions have ever been met in any real white dwarf stars.

 \bfig \centerline{\psfig{file=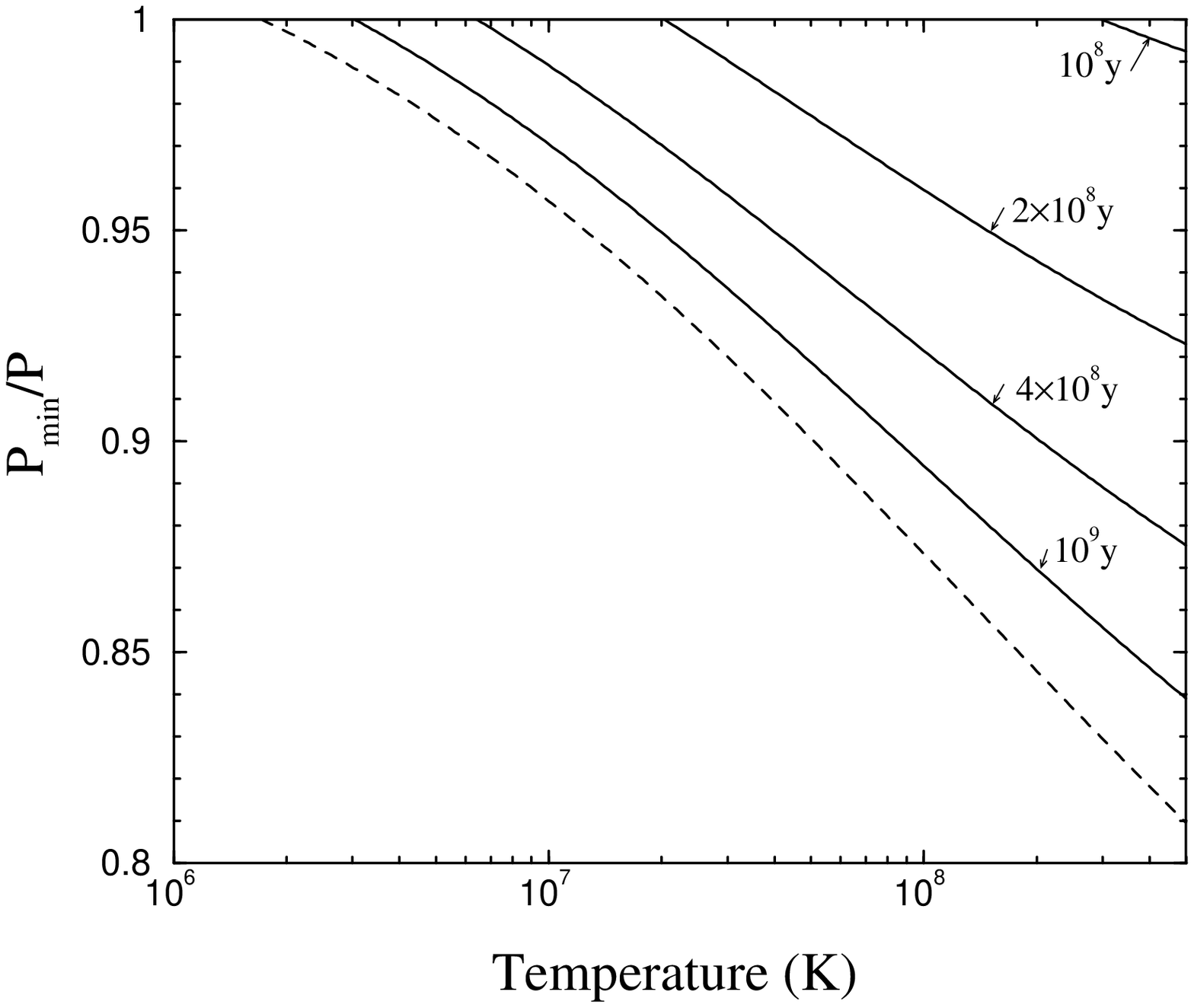,height=2.4in}} \vskip 0.3cm
\caption{Rotation periods $P$ (as a function of temperature) for
1.3$M_\odot$ white dwarf stars where the growth time of the $m=2$
$r$-mode has the prescribed value.  The dashed curve is the critical
rotation period $P_c$ where the growth time is infinite.
\label{fig5}} \efig

\acknowledgements

I thank L.~Bildsten, J.-P.~Lasota, Y.~Levin, S.~Phinny and R.~Wagoner for
helpful conversations concerning this work.  This work was supported
by NSF grant PHY-9796079 and NASA grant NAG5-4093.

\vfill\eject

\end{document}